# Doping effects on the valence bond solid of $Li_2RuO_3$ with Mn substitution

Seokhwan Yun[1,2,3], Ki Hoon Lee[3,4], Chaebin Kim[1,2,3], Junghwan Park[5], Min-Gyu Kim[6], Deok-Yong Cho[7], D. I. Khomskii[8], and Je-Geun Park[1,2,3*]

[1]*Center for Quantum Materials, Seoul National University, Seoul 08826, Korea*

[2]*Department of Physics and Astronomy, Seoul National University, Seoul 08826, Korea*

[3]*Center for Correlated Electron Systems, Institute for Basic Science, Seoul 08826, Korea*

[4]*Department of Physics, Incheon National University, Incheon 22012, Korea*

[5]*Samsung SDI Co. Ltd., Suwon, Gyeonggi, 16678, Korea*

[6]*Beamline Research Division, Pohang Accelerator Laboratory (PAL), Pohang 790-784, Korea*

[7]*IPIT and Department of Physics, Jeonbuk National University, Jeonju 54896, Korea*

[8]*II. Physikalisches Institut, Universität zu Köln, 50937 Köln, Germany*

**Corresponding author: jgpark10@snu.ac.kr**

**Abstract:** $Li_2RuO_3$ with a honeycomb structure undergoes a drastic transition from a regular honeycomb lattice with the C2/m space group to a valence bond solid state of the $P2_1/m$ space group with an extremely strong dimerization at 550 K. We synthesized $Li_2Ru_{1-x}Mn_xO_3$ with a full solid solution and investigated doping effects on the valence bond solid state as a function of Mn content. The valence bond solid state is found to be stable up to x = 0.2, based on our extensive experiments: structural studies, resistivity, and magnetic susceptibility. On the other hand, the extended x-ray absorption fine structure analyses show that the dimer local structure remains robust even above x = 0.2 with a minimal effect on the dimer bond length. This indicates that the locally-disordered dimer structure survives well into the Mn-rich phase even though the thermodynamically stable average structure has the C2/m space group. Our results prove that the dimer formation in $Li_2RuO_3$ is predominantly a local phenomenon driven by the formation of orbitally-assisted metal-metal bonds and that these dimers are relatively robust against doping-induced disorder.

## 1. Introduction

For most transition metal oxides, strong covalent bonding between *d*-orbitals of transition metal ions with oxygen p orbitals is essential to describe the various properties found in the materials [1]. Although not so common, there are also situations where a direct overlap between *d*-orbitals plays a significant role. In this case, various other factors have become increasingly important. And they need to be considered with care, such as the position of a transition metal in a periodic table, the shape of the wave functions in the $t_{2g}$ manifold, and the geometry of the network of metal-ligand polyhedra [2]. Under certain conditions, the transition metal ions can also form well-defined clusters, and the electronic wave function of these clusters could then be described using a molecular orbital picture. Such metal clusters in the solid form a periodic array called a valence bond solid (VBS) or, sometimes, a valence bond crystal [3–5]. And the system's periodicity and pattern of those systems are related to the cluster's internal degrees of freedom. Not surprisingly, the orbital degrees of freedom are essential for such clusters' orbital-selective behavior [6–8] with a well-known example of $Tl_2Ru_2O_7$ [9].

All these features are now seen in the example of $Li_2RuO_3$. It has a layered honeycomb structure with Ru-O layers separated by the Li layer, and the honeycomb Ru layers are composed of edge-sharing $RuO_6$ octahedra. Miura et al. [10] reported a peculiar structural transition in this system, with strong Ru dimers forming a herringbone pattern in the honeycomb layer. The dimerized Ru-Ru bond length was 2.57 Å at 300 K [10] – much shorter than the Ru-Ru metallic bonds of 2.65 Å. On the other hand, the other Ru-Ru bonds between dimers (inter-dimer bonds) of about 3.05 Å are much longer than an intra-dimer bond of about 2.57 Å. This difference between two Ru-Ru bonds is the largest ever reported so far in Ru compounds. Another interesting point is that this dimerized system goes through a structural transition at an exceptionally high transition temperature of Tc= 550 K, accompanied by a concomitant change in the space group from $P2_1/m$ below Tc to C2/m above. It was also pointed out based on the Wilson ratio analysis that the dimerized phase exhibits unusual correlation effects [11].

According to the previous X-ray diffraction data, Ru dimerization on the average disappears above Tc [12]. On the other hand, a different picture emerges out of the pair distribution function analysis (PDF) with the total scattering measurement that the dimers still survive well above the transition temperature on a local scale [12]. These particular results call for a revision of the simple dimer formation picture at the transition temperature. Instead, it implies that one has to visualize this average C2/m structure containing random and fluctuating Ru dimers. This also agrees that the system exhibits a diminished local spin moment of $S$ = 1/2, instead of the $S$ = 1 expected for the typical $t_{2g}^4$ electron configuration of $Ru^{4+}$ [12]. With these thermally fluctuating dimers, this phase may be called a valence bond liquid (VBL) state.

The present study aims to investigate the doping effect on the dimerized state by replacing Ru with Mn. As $Li_2MnO_3$ has the same honeycomb structure as the C2/m space group [13], one could expect that it would form a solid solution with $Li_2RuO_3$ all across the full doping range. A small difference in the unit cell volumes of 5% between the two end compounds is also a favorable factor for forming the full range of solid solutions, with the possibility of fine-tuning and of precise control over the dimer phase. These expectations are indeed confirmed experimentally: we managed to synthesize $Li_2Ru_{1-x}Mn_xO_3$ in the full concentration range $0 \leq x \leq 1$. In this study, we found that the thermodynamically stable VBS phase is only confined up to 20% of Mn doping, above which the transition becomes invisible by all the thermodynamic and transport measurements we carried out. However, our local structural study using Extended X-ray absorption fine structure (EXAFS) demonstrates that this VBS state exists locally and probably becomes VBL for the Mn-rich region. Therefore, our Mn doping plays a similar role as the temperature across the unique phase transition, thereby offering an exciting novel window into this intriguing physics.

## 2. Experiments

We prepared 12 different samples for this study altogether. Polycrystalline samples of $Li_2Ru_{1-x}Mn_xO_3$ (x = 0, 0.03, 0.05, 0.08, 0.1, 0.2, 0.4, 0.5, 0.6, 0.8, 0.9 and 0.95) were synthesized by a solid-state reaction method. The starting materials were $Li_2CO_3$ (99.995%, Alfa Aesar), $RuO_2$ (99.95%, Alfa Aesar), and $MnO_2$ (99.995%, Alfa Aesar). We first dried the starting materials at 600 K for 6 h due to their hygroscopic character. The stoichiometric quantity of each compound plus 5% excess of $Li_2CO_3$ was placed in an alumina crucible, and the mixture sintered sequentially at 700 and 900 °C for 12 h at each temperature. After that, each mixture was pelletized and heated at 1000 °C for 24 h. The samples' structures were confirmed by powder X-ray diffraction (XRD) using a Rigaku Miniflex2 (Cu target, suppressing $K_\beta$ with Ni-filter). The lattice parameters of each sample were refined with the Le Bail method.

We measured a high temperature (HT) resistivity of each pelletized sample with the four-probe method using our home-built setup. The voltage difference between $I^+/I^-$ electrodes was kept below 0.2 V to prevent any possible charging effects from the highly mobile $Li^+$ ions [14]. The HT magnetic susceptibility measurements were carried out using a magnetic property measurement system (MPMS-3, Quantum Design). We sealed the sample with non-magnetic zirconium cement during the measurement to improve the samples' thermal conduction. We also measured the enthalpy change across the phase transition using differential scanning calorimetry (Discovery DSC, TA Instrument). The measurement was carried out with a heating rate of 10 K/min under an $N_2$ environment.

To examine the local structure of heavily doped samples, we employed the Extended X-ray absorption fine structure. It is a technique specially designed to probe local structure around a particular ion [15]. The EXAFS spectra of $Li_2Ru_{1-x}Mn_xO_3$ were measured at the Ru *K*-edge in a transmission mode at the beamline 10C at Pohang Light Source (PLS), Korea. The samples were sealed in polyethylene for the room temperature measurement. For the HT measurement, the sample is mixed with BN at a 1:1 weight ratio and pelletized. The data were processed and analyzed with Demeter [16]. The fitting curves and the summary of the fitting parameters are given in Figure 6 and Table 1.

## 3. Result & Analysis

$Li_2RuO_3$ ($P2_1/m$) and $Li_2MnO_3$ (C2/m) form a layered honeycomb lattice with similar crystal structures, as shown in Figure 1a [10,13]. Both structures are composed of edge-sharing octahedra, but only $Li_2RuO_3$ has the contracted transition metal bonds with the space group $P2_1/m$ at room temperature. The XRD data of the solid solution in Figure 1b shows that with increasing Mn composition, two peaks at 44° and 45° come closer and almost merge at x = 0.2. It signals that the structure with $P2_1/m$ is no longer stable for $x \geq 0.2$, at least as an average structure. According to our X-ray data analysis, the system has the C2/m structure of $Li_2MnO_3$ for $x \geq 0.2$. The lattice parameters of each sample were refined with the Le Bail method [17]. The unit cell volume of the $Li_2Ru_{1-x}Mn_xO_3$ solid solution in Figure 2a decreased monotonically with increasing x.

The blue-green and burgundy lines in Figure 2a are linear fitting results using Vegard's law for the two different regions: one is for $0 \leq x \leq 0.2$ and another for $0.4 \leq x \leq 1$. These linear fits show two distinct regions in terms of the fitting: one (burgundy) for the smaller doping is that of the VBS while the other (blue-green) for the higher doping range, for the phase with the C2/m space group. Another interesting observation is the breakdown of Vegard's law, which is rather rarely seen. That we have observed such a rare breakdown of Vegard's law by Mn-doping for x larger than 0.2 must be related to the nature of the dimer structure and its doping effect. For example, although the dimer phase is seen to be stable only up to x = 0.2, the decreasing yet persisting gap to x = 0.4 or higher between the two straight lines implies that the dimer phase may well survive locally even if it is no longer thermodynamically stable. This point will be further investigated by our EXAFS experiments to be discussed later in the paper.

Dimerization for $x \leq 0.2$ is also reflected in the lattice parameters' doping dependence shown in Fig. 2b. On the other hand, the inter-layer spacing $c \cdot \sin\beta$ shows the typical behavior of decreasing linearly with increasing Mn because Mn has a smaller ionic radius than Ru ($Mn^{4+}$: 0.53 Å / $Ru^{4+}$: 0.62 Å) [18]. A more drastic observation is that the distortion parameter $u$ $(= b/(a\sqrt{3}) - 1)$ , shown in Figure 2c, quantifying the dimer distortion of the $P2_1/m$ phase, exhibits an apparent suppression with Mn doping before disappearing for x > 0.2, consistent with our conclusion above. Therefore, our XRD data confirm that the solid solution in the $P2_1/m$ phase has an additional volume reduction related to the distortion in the honeycomb layer, and this average distortion exists up to x = 0.2.

We also studied how the physical properties of the system evolve upon Mn doping. The curves in Figure 3a are the normalized magnetic susceptibility measured with a magnetic field of 1 T. There are hysteresis loops in the curves for $x \leq 0.2$ due to the phase transition. Interestingly, both physical properties have anomalies at temperatures considering those in pure $Li_2RuO_3$ [10], but they deviate from each other with increasing Mn. For instance, the magnetic transition temperatures rose from 530 K for x = 0 to 560 K for x = 0.03. On the other hand, the resistivity transition temperature decreased monotonically with increasing x (Figure 3b, c). The temperature difference for x = 0.15 was about 50 K. Previous studies by Mehlawat and Ponosov also reported that the transition of $Li_2RuO_3$ is a combination of two consecutive phase transitions [19,20], and they have both the nature of the first- and second-order phase transitions. Our result shows that the transition is indeed complicated and seems to reveal such features more readily upon doping. Given the results of electric and magnetic properties presented above, one sees that the phase diagram of $Li_2Ru_{1-x}Mn_xO_3$ systems can be divided into the $P2_1/m$ and the C2/m phases, and the solid solution in the $P2_1/m$ phase has a phase transition behaving like that in the pure $Li_2RuO_3$. Another point of note is the gradual increase of the resistivity value at 600 K with Mn doping. It implies that the tiny charge gap of $Li_2RuO_3$ [11] gets considerably increased by Mn doping, pushing it towards a Mott insulating regime.

Figure 4a is the $k^3$-weighted Fourier transform (FT) of the EXAFS spectra taken at 300 K at the Ru *K*-edge for $Li_2Ru_{1-x}Mn_xO_3$ (x = 0, 0.05, 0.1, 0.2 and 0.4, the range of FT: 3 ~ 14 $Å^{-1}$). A peak related to each scattering path on the spectra is generally 0.3 ~ 0.4 Å shorter than the actual interatomic length because of the phase shift by the potentials near scattering and absorbing ion. In the previous research, the peaks around 1.5 and 2.2 Å were identified as single scattering paths for Ru-O (2.0 ~ 2.1 Å) and Ru-Ru (Dimer, 2.57 Å), respectively. And the EXAFS spectra at Mn *K*-edge show that Mn does not form dimers [21]. Our 300 K data show that the Ru dimer's length is not affected by Mn doping regardless of the system; the 2.2 Å peak is not shifted by Mn doping up to x = 0.4, where the system has the space group of C2/m (Figure 4b).

Figure 4c is the temperature dependence of the EXAFS spectra for $Li_2Ru_{0.9}Mn_{0.1}O_3$. The peak at 2.2 Å, which represents the single scattering path between Ru ions in the dimer, is slightly shifted with increasing temperature (Figure 4d). However, its intensity decreases because of thermal broadenings. Besides, there is no pronounced change at the transition, unlike the EXAFS results on other clustered systems such as $VO_2$ and 1*T*-$CrSe_2$ [22,23]. This analysis shows that the Ru ion's surroundings are not changed much, and the dimers still exist above the phase transition temperature, at least as a local structure.

To examine the thermodynamic nature of the transition with doping more directly, we measured the transformation heat of $Li_2Ru_{1-x}MnxO_3$ solid solution in the $P2_1/m$ phase during the phase transition (Figure 5a). The phase transition temperature linearly decreases with doping, and its value is close to that deduced from the resistivity data in Figure 3c. The deviation between the two sets of transition temperatures, seen in Fig.3, is due to the first-order phase transition hysteresis. The integrated area for x = 0 is found to be 1.0 kW·K/mol, and the enthalpy change is 6.0 kJ/mol (Heating rate: 10 K/min) during the phase transition. This value corresponds to 62 meV per chemical formula (Figure 5b). A previous study by Kimber et al. [12] reported that the calculated energy difference between the Armchair ($P2_1/m$) and Parallel (C2/m) structure is 42 meV, which is suggested to represent the energy difference between the VBS and VBL phases. Our result is bigger by 20 meV than their value, which we think is not due to the volume change: the 1 $Å^3$ (= $10^{-30}$ $m^3$) unit cell volume variation only contributed about $6.0\times10^{-4}$ meV to the enthalpy at normal atmospheric pressure. Instead, we believe that it has an electronic origin. $Li_2RuO_3$ is an insulator over the entire temperature range. Still, the calculated electronic density of states in Kimber's study has no electronic energy gap near the Fermi energy, regardless of the phase transition. This discrepancy might as well lead to the underestimation of the electronic energy reduction by the phase transition.

According to our analysis, the enthalpy change $\Delta H$ of the phase transition decreases as $x$ increases. This indicates that Mn's substitution for Ru gradually breaks the dimers consistent with other data discussed above (Figure 5b). Furthermore, it reduces the entropy variation per Ru ion $\Delta S$ during the phase transition: the linear fit shows that replacing Ru ion by one Mn ion reduces the $\Delta S$ by 3.51 ($\partial(\Delta S)/\partial x|_{x=0}$ = -3.51). The $\Delta S$ per ruthenium ion of pure $Li_2RuO_3$ is 1.3, closer to one-third of $\partial(\Delta S)/\partial x|_{x=0}$. But in the case of x = 0.2, in the vicinity of the boundary of the $P2_1/m$ and the C2/m phases, the averaged entropy variation is much reduced only to 0.62. Again, this observation paints a picture that the dimer formation is gradually suppressed by Mn doping. Interestingly, the previous theoretical discussion demonstrated that the orbital degeneracy causes spontaneous dimerization of spins and induces VBS's herringbone pattern [24].

## 4. Discussion

First of all, we would like to note that before our work, there have been some reports of the doping of $Li_2RuO_3$; see, for example, the results for Ti-doped $Li_2RuO_3$ [14], $Li_2Ru_{0.5}Mn_{0.5}O_3$ [21,25], and Ir-doped $Li_2RuO_3$ [26]. What distinguishes our work from the previous studies is that we have made a comprehensive study over the entire doping range. And we have also used experimental techniques covering both global and local structures in addition to the thermodynamics tools like the heat of the transformation or bulk magnetic and transport properties. Using this extensive study, we could paint a complete picture of the doping effect on the dimer structure of $Li_2RuO_3$.

The experimental results of the structural deformation, resistivity, and magnetic susceptibility showed that the ordered dimer phase with $P2_1/m$ remains robust up to x = 0.2. Our EXAFS data for x = 0.4, on the other hand, show that the Ru dimer still exists locally in the C2/m compounds. As our XRD data do not show any sign of the dimer phase, we conclude that both the highly-doped and high-temperature phases contain disordered and possibly mobile dimers. It is also confirmed by the observed deviations from Vegard's law.

The previous theoretical study demonstrated that the orbital degeneracy is instrumental in the spontaneous dimerization of $Li_2RuO_3$ and induces the VBS's herringbone pattern [24]. This study also mentioned other possible types of orbital patterns, such as the opened or closed chains. Although those patterns were rejected in the perfect Ru honeycomb system, such states could contribute to the proper ground state in the heavily doped system because the Mn substitution breaks the dimers and makes the Ru network finite. To verify this scenario, further calculations are required, such as, e.g., the Monte Carlo simulations. For higher doping than x=0.2, we would like to note that these more Mn-rich samples exhibit the locally surviving dimer phase in our EXAFS results. Simultaneously, there are visible signs of difference in the transition temperature between the resistivity and susceptibility. Taken together, it points to a picture that the system is most likely to be phase-separated and inhomogeneous, consisting of regions with different properties/structures.

Another point worth noting is that the VBS phase requires two features: the metal-insulator transition and the singlet formation. Intriguingly, these two can occur at precisely the same temperatures in pure $Li_2RuO_3$. It implies that the two entities are coupled to one another: they are triggered by the same source of orbitals in any case. However, in Mn doping, they appear to occur at slightly different temperatures, as seen in the resistivity and the susceptibility (see Fig. 3). Interestingly, the DSC results in Fig. 5 exhibit the transition at the same temperature as the resistivity. We can identify the transition seen by both resistivity and DSC with the metal-insulator transition while the one in the susceptibility to the singlet formation. Not surprisingly, the metal-insulator transition accounts for most of the entropy change involved in the dimer formation. These observations strongly indicate that the two in-principle independent mechanisms of the VBS phase: the metal-insulator transition and the singlet formation, may well be split by Mn doping, which is new insight and very revealing for us to grasp a better understanding of the drastic dimer formation in $Li_2RuO_3$.

Finally, the thermal analysis result shows that the phase transition's enthalpy change was underestimated in the previous DFT calculation. In the calculation results, the band structures below and above the phase transition have a non-zero electronic density of states at the Fermi energy [12]. Our previous research verified that the electronic correlation effects are critical to account for the physical properties of $Li_2RuO_3$, including the anisotropy [27]. The mismatch between our experimental and the calculation results could originate from the correlations [27].

In summary, the structural deformation, resistivity, and magnetic susceptibility of $Li_2Ru_{1-x}Mn_xO_3$ show that the valence bond solid phase maintains up to x = 0.2. But the local structure study with EXAFS indicates that the dimers still exist above x = 0.2, and the Mn substitution does not influence the dimer's bond length. These results and the comparison with the other doping studies indicate that the dimerization in $Li_2RuO_3$ is mostly a local phenomenon, causing the formation of metal-metal bonds, primarily promoted by the orbital degrees of freedoms.

Dimers in $Li_2RuO_3$ are relatively robust, both to temperature increase and strong doping, which speaks for their local character. One can think that the dimerization, or more generally, the formation of molecular-like clusters, often observed in correlated solids, see, e.g. [28], should also have similar features.

**Acknowledgments**

Work at the Center for Quantum Materials was supported by the Leading Researcher Program of the National Research Foundation of Korea (Grant No. 2020R1A3B2079375). The work of D.Kh. was supported by the Deut sche Forschungsgemeinschaft (DFG, German Research Foundation) - Project number 277146847 - CRC 1238.

**References**

[1] D. I. Khomskii, *Transition Metal Compounds* (Cambridge University Press, Cambridge, 2014).

[2] D. I. Khomskii and S. V Streltsov, Chem. Rev. 0c00579 (2020).

[3] M. Ye, H. S. Kim, J. W. Kim, C. J. Won, K. Haule, D. Vanderbilt, S. W. Cheong, and G. Blumberg, Phys. Rev. B **98**, 201105 (2018).

[4] Y. Haraguchi, C. Michioka, M. Ishikawa, Y. Nakano, H. Yamochi, H. Ueda, and K. Yoshimura, Inorg. Chem. **56**, 3483 (2017).

[5] J. P. Sheckelton, J. R. Neilson, D. G. Soltan, and T. M. McQueen, Nat. Mater. **11**, 493 (2012).

[6] S. V. Streltsov and D. I. Khomskii, Phys. - Uspekhi **60**, 1121 (2017).

[7] M. A. McGuire, J. Yan, P. Lampen-Kelley, A. F. May, V. R. Cooper, L. Lindsay, A. Puretzky, L. Liang, S. Kc, E. Cakmak, S. Calder, and B. C. Sales, Phys. Rev. Mater. **1**, 1 (2017).

[8] M. W. Haverkort, Spin and Orbital Degrees of Freedom in Transition Metal Oxides and Oxide Thin Films Studied by Soft X-Ray Absorption Spectroscopy, 2005.

[9] S. Lee, J.-G. Park, D. T. Adroja, D. Khomskii, S. Streltsov, K. A. McEwen, H. Sakai, K. Yoshimura, V. I. Anisimov, D. Mori, R. Kanno, and R. Ibberson, Nat. Mater. **5**, 471 (2006).

[10] Y. Miura, Y. Yasui, M. Sato, N. Igawa, and K. Kakurai, J. Phys. Soc. Japan **76**, 033705 (2007).

[11] J. Park, T.-Y. Tan, D. T. Adroja, A. Daoud-Aladine, S. Choi, D.-Y. Cho, S.-H. Lee, J. Kim, H. Sim, T. Morioka, H. Nojiri, V. V. Krishnamurthy, P. Manuel, M. R. Lees, S. V. Streltsov, D. I. Khomskii, and J.-G. Park, Sci. Rep. **6**, 25238 (2016).

[12] S. A. J. Kimber, I. I. Mazin, J. Shen, H. O. Jeschke, S. V. Streltsov, D. N. Argyriou, R. Valentí, and D. I. Khomskii, Phys. Rev. B **89**, 081408(R) (2014).

[13] S. Lee, S. Choi, J. Kim, H. Sim, C. Won, S. Lee, S. A. Kim, N. Hur, and J. G. Park, J. Phys. Condens. Matter **24**, (2012).

[14] M. Sathiya, A. M. Abakumov, D. Foix, G. Rousse, K. Ramesha, M. Saubanère, M. L. Doublet, H.

Vezin, C. P. Laisa, A. S. Prakash, D. Gonbeau, G. Vantendeloo, and J. M. Tarascon, Nat. Mater. **14**, 230 (2015).

[15] J. J. Rehr and R. C. Albers, Rev. Mod. Phys. **72**, 621 (2000).

[16] B. Ravel and M. Newville, in *J. Synchrotron Radiat.* (International Union of Crystallography, 2005), pp. 537–541.

[17] A. Le Bail, Powder Diffr. **20**, 316 (2005).

[18] R. D. Shannon, Acta Crystallogr. Sect. A **32**, 751 (1976).

[19] Y. S. Ponosov, E. V Komleva, and S. V Streltsov, Phys. Rev. B **100**, 134310 (2019).

[20] K. Mehlawat and Y. Singh, Phys. Rev. B **95**, 075105 (2017).

[21] Y. Lyu, E. Hu, D. Xiao, Y. Wang, X. Yu, G. Xu, S. N. Ehrlich, K. Amine, L. Gu, X. Q. Yang, and H. Li, Chem. Mater. **29**, 9053 (2017).

[22] T. Yao, X. Zhang, Z. Sun, S. Liu, Y. Huang, Y. Xie, C. Wu, X. Yuan, W. Zhang, Z. Wu, G. Pan, F. Hu, L. Wu, Q. Liu, and S. Wei, Phys. Rev. Lett. **105**, 226405 (2010).

[23] S. Kobayashi, N. Katayama, T. Manjo, H. Ueda, C. Michioka, J. Sugiyama, Y. Sassa, O. K. Forslund, M. Månsson, K. Yoshimura, and H. Sawa, Inorg. Chem. **58**, 14304 (2019).

[24] G. Jackeli and D. I. Khomskii, Phys. Rev. Lett. **100**, 147203 (2008).

[25] D. Mori, H. Kobayashi, T. Okumura, H. Nitani, M. Ogawa, and Y. Inaguma, Solid State Ionics **285**, 66 (2016).

[26] H. Lei, W.-G. Yin, Z. Zhong, and H. Hosono, Phys. Rev. B **89**, 020409 (2014).

[27] S. Yun, K. H. Lee, S. Y. Park, T. Y. Tan, J. Park, S. Kang, D. I. Khomskii, Y. Jo, and J. G. Park, Phys. Rev. B **100**, 165119 (2019).

[28] S. V. Streltsov and D. I. Khomskii, Proc. Natl. Acad. Sci. **113**, 10491 (2016).

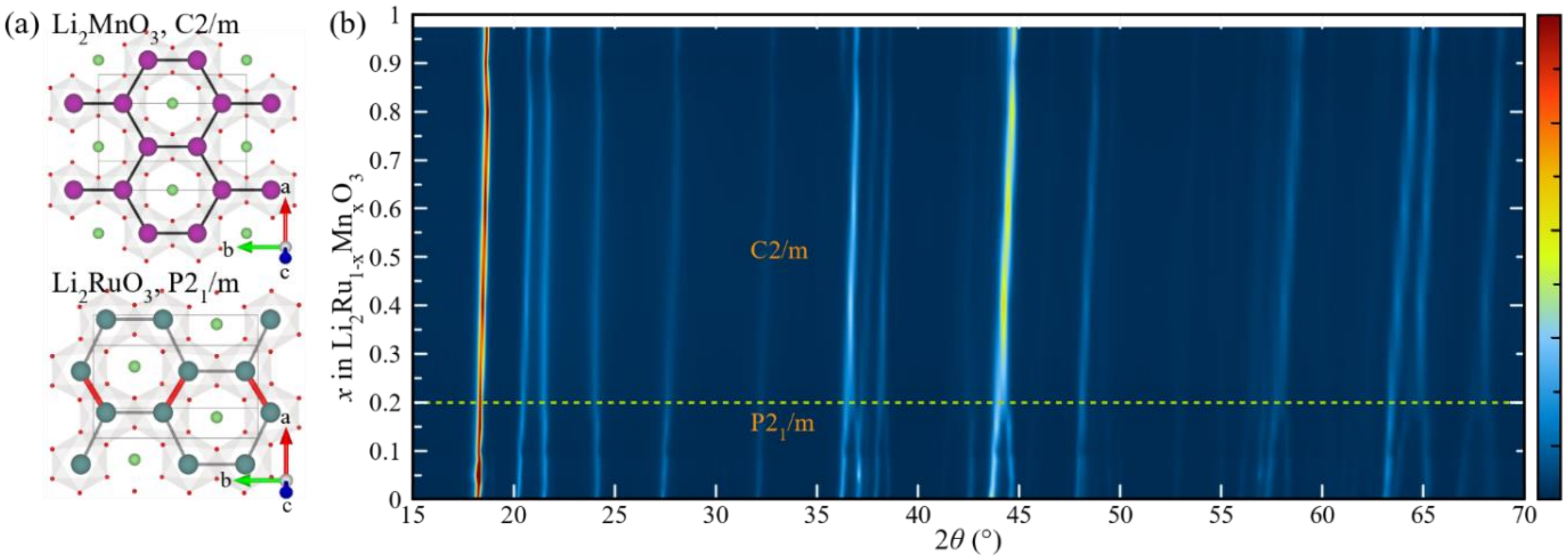

**Figure 1** (a) The crystal structures of $Li_2MnO_3$ (top) and $Li_2RuO_3$ (bottom). Both have a layered honeycomb structure separated by $Li^+$ ions, but only $Li_2RuO_3$ has strong dimerization. The dimer (red) bonds are 2.57 Å, while the non-dimer (black) bonds are about 3.05 Å. In contrast, $Li_2MnO_3$ has regular inter-transition metal ion bonds in the range of 2.82 ~ 2.84 Å. (b) XRD data for the $Li_2Ru_{1-x}Mn_xO_3$ systems with the P $2_1$/m space group for $x \leq 0.2$ and the C 2/m space group for $x > 0.2$.

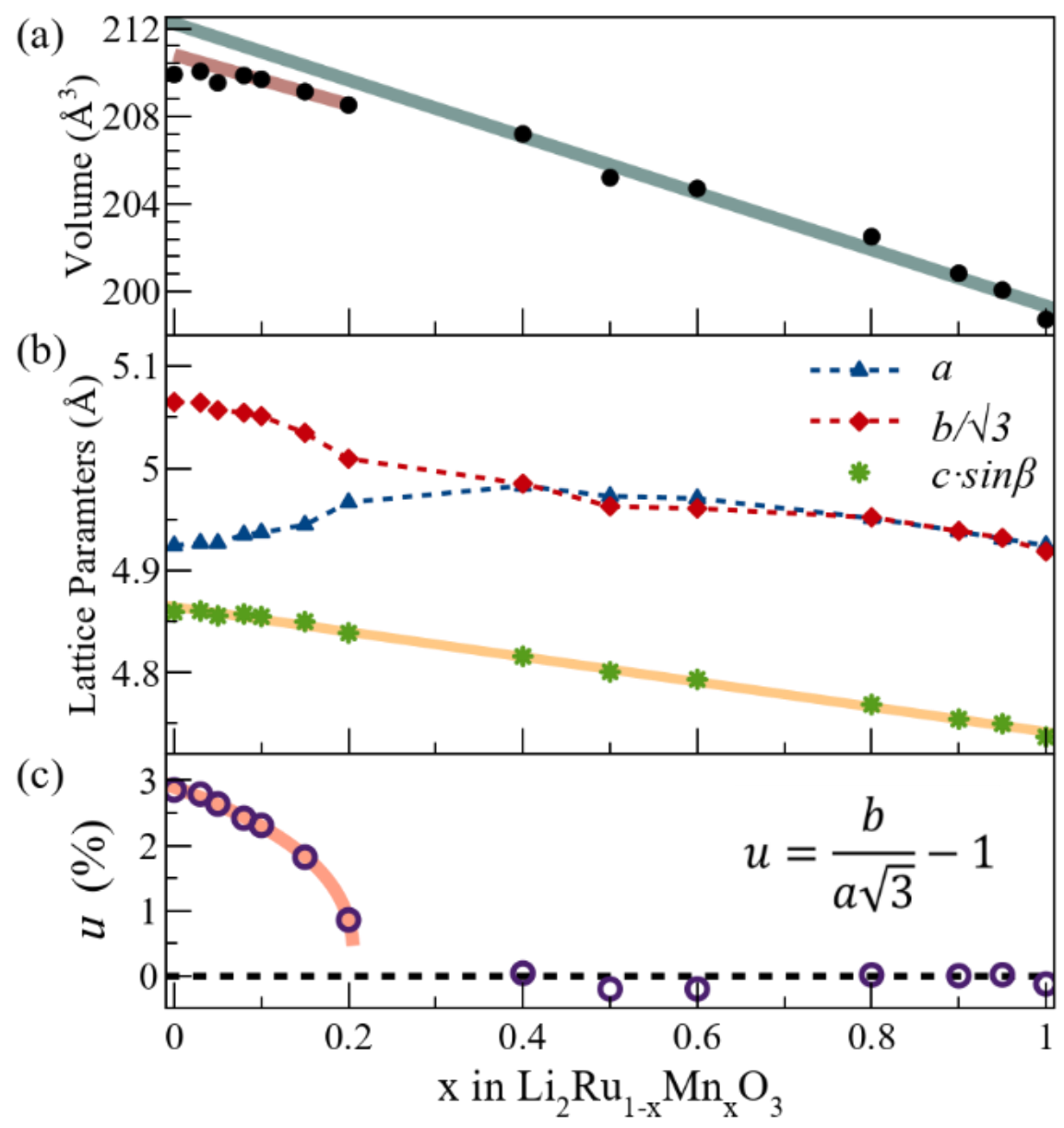

**Figure 2** (a) Unit cell volumes of the $Li_2Ru_{1-x}Mn_xO_3$ systems refined by the Le bail method. The blue-green line is a fitting result for the systems' volume data with x > 0.2, and the burgundy line is that of the systems with $0 \leq x \leq 0.2$. (b) The lattice parameters *a* (blue triangles), *b* (red diamonds), and the interlayer distance ($c$·sin$\beta$, green snowflake) of $Li_2Ru_{1-x}Mn_xO_3$ systems. Both *b* and $c$·sin$\beta$ decrease monotonically with increasing Mn doping, while *a* maximum is about at Mn 40 %. The orange line in (b) is a guide to the eyes. (c) The distortion parameter $u = \frac{b}{a\sqrt{3}} - 1$ is plotted as a function of Mn doping.

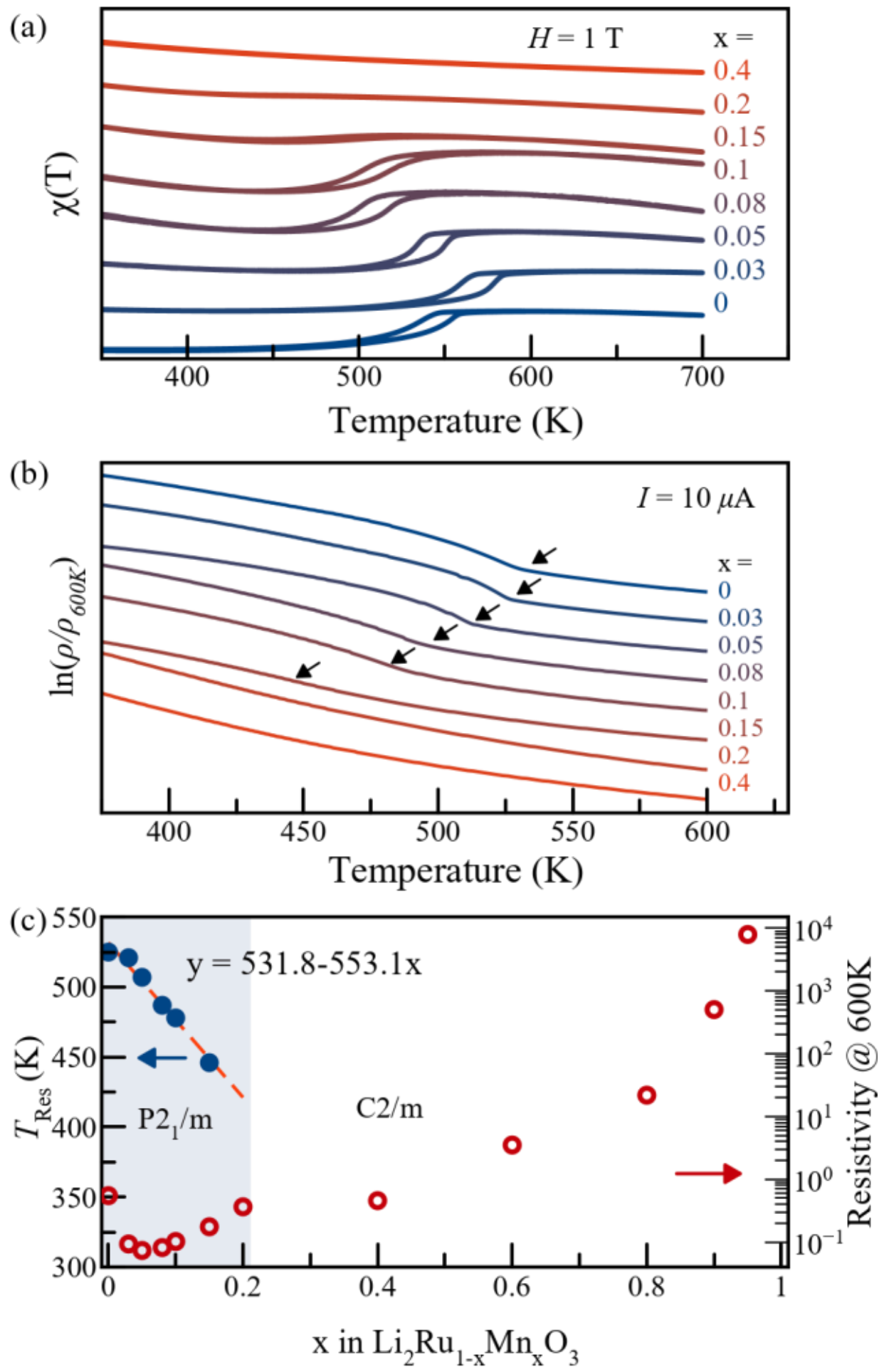

**Figure 3** (a) Normalized magnetic susceptibility data taken at $H$ = 1 T for $Li_2Ru_{1-x}Mn_xO_3$ systems. (b) Normalized resistivity data with $I$ = 10 $\mu$A for $Li_2Ru_{1-x}Mn_xO_3$ systems. The black arrows indicate the phase transition temperature of each resistivity curve. (c) Resistivity phase transition temperature ($T_{Res}$, left) and the $Li_2Ru_{1-x}Mn_xO_3$ systems' resistivity at 600 K (right). The red dashed line and the equation are a line fitting result of the transition temperatures.

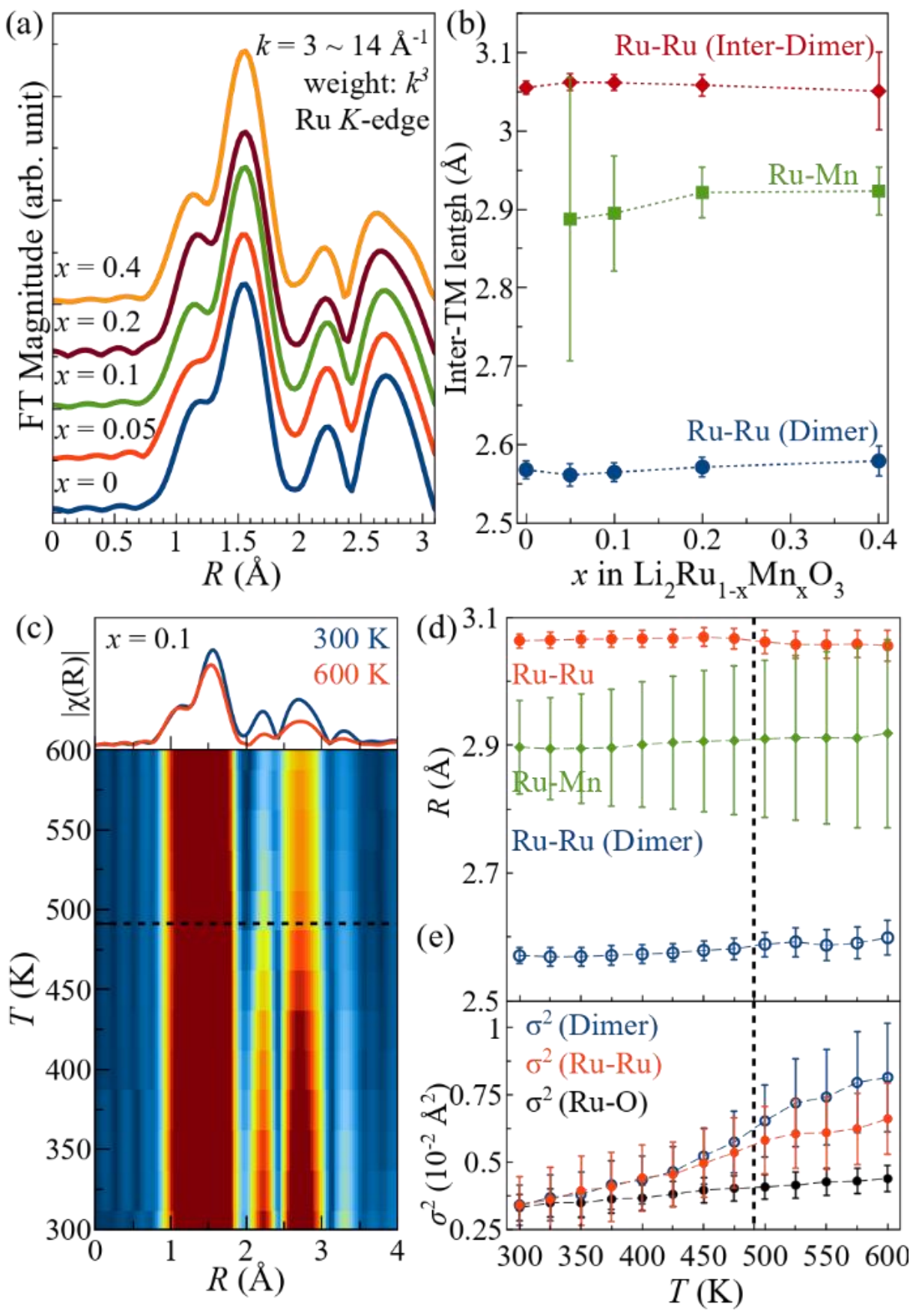

**Figure 4** (a) The $k^3$-weighted Fourier transform magnitudes of the Ru $K$-edge EXAFS spectra of the $Li_2Ru_{1-x}Mn_xO_3$ (x = 0, 0.05, 0.1, 0.2, 0.4) systems. (b) The doping dependency of the inter-transition metal distances of Ru and Mn in the honeycomb layer. The blue circles (red diamond) indicate Ru-dimer's lengths (inter-dimer), and the green squares indicate the distance between Ru and Mn. (c) Temperature dependence of the EXAFS spectra of $Li_2Ru_{0.9}Mn_{0.1}O_3$. The temperature dependencies of (d) the distances between Ru and Mn in the honeycomb layer and (e) their thermal factors. The black dash lines in all graphs are the structural phase transition temperatures of the system.

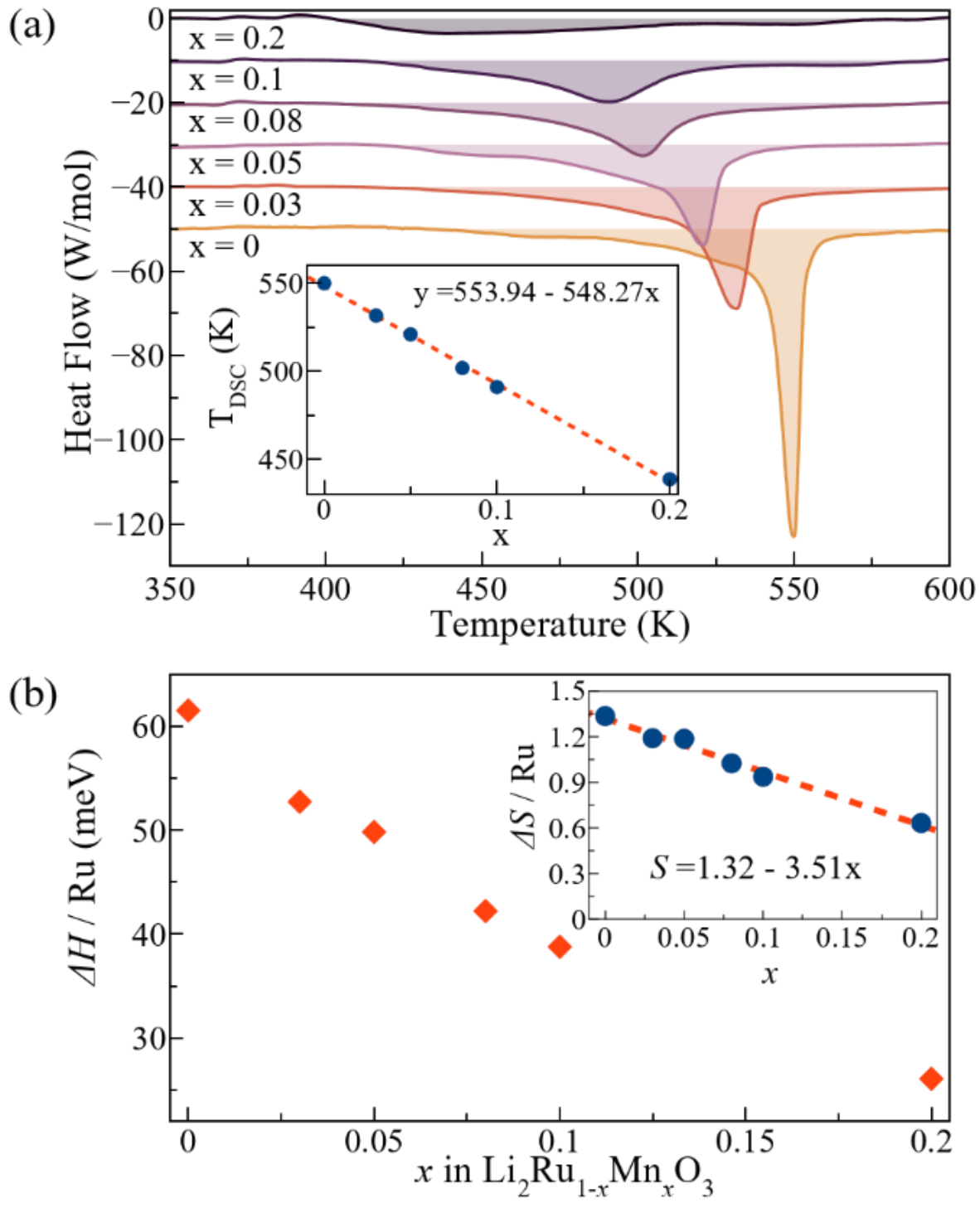

**Figure 5** (a) Differential scanning calorimetry (DSC) heat flow curves for a series of $Li_2Ru_{1-x}Mn_xO_3$ systems. The inset graph shows the phase transition temperature ($T_{DSC}$) of the systems. The linear fitting result is shown as a dashed red line. The heating rate is 10 K/min. (b) Variation of enthalpy change $\Delta H$ per Ru ion with x for $Li_2Ru_{1-x}MnO_3$. The inset graph shows the calculated entropy change $\Delta S\,(=\int dQ/T)$ per Ru ion during the phase transition.

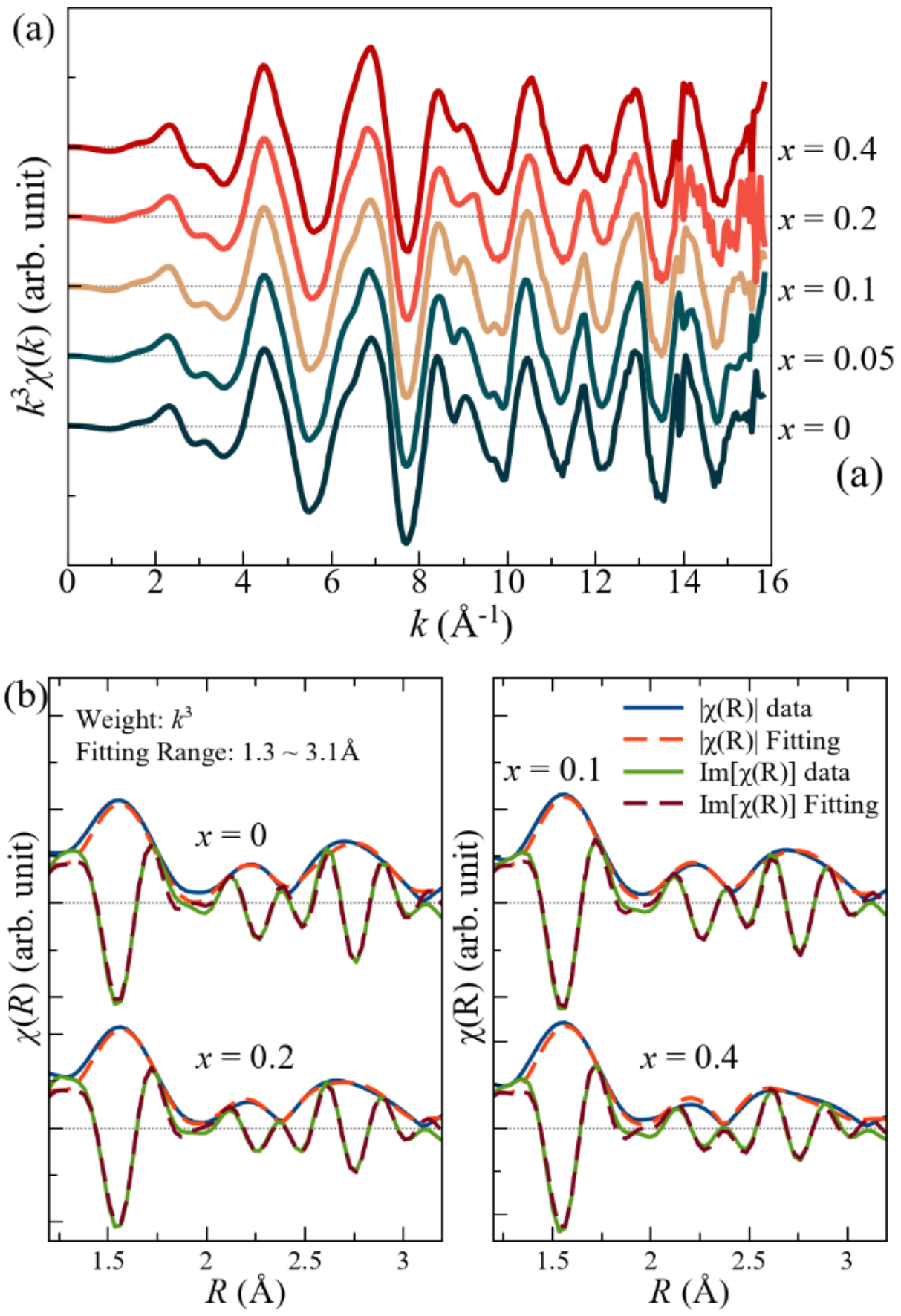

**Figure 6** (a). $k^3$-weighted $\chi(k)$ and Fitting data and curves of EXAFS signals for $Li_2Ru_{1-x}Mn_xO_3$ (x = 0, 0.05, 0.1, 0.2 and 0.4)

**Table 1**. Fitting information of EXAFS signals for $Li_2Ru_{1-x}Mn_xO_3$ (x = 0, 0.05, 0.1, 0.2 and 0.4).

| | | $x$ = 0 | $x$ = 0.05 | $x$ = 0.1 | $x$ = 0.2 | $x$ = 0.4 |
|---|---|---|---|---|---|---|
| Input | $N_{\mathrm{O,\ Ru\text{-}O}}$ | 6 | 6 | 6 | 6 | 6 |
| | $N_{\mathrm{Ru(1),\ Dimer}}$ | 1 | 1 | 1 | 1 | 1 |
| | $N_{\mathrm{Ru(2),\ Inter\text{-}Dimer}}$ | 2 | 1.8 | 1.6 | 1.2 | 0.4 |
| | $N_{\mathrm{Mn,\ Ru\text{-}Mn}}$ | - | 0.2 | 0.4 | 0.8 | 1.6 |
| | # of variables | 7 | 9 | 9 | 9 | 9 |
| Output | $R$-Factor | 0.020657 | 0.01972 | 0.012115 | 0.011302 | 0.024187 |
| | $\Delta E_0$ | -1.591 ± 2.224 | -0.878 ± 2.550 | -0.743 ± 1.903 | 0.598 ± 1.716 | 0.619 ± 2.570 |
| Correlations | $\Delta E_0$ & $\Delta R_{Ru\text{-}\mathrm{O}}$ | 0.8177 | 0.8436 | 0.8408 | 0.8383 | 0.8477 |
| | $\Delta E_0$ & $\Delta R_{\mathrm{Dimer}}$ | 0.4739 | 0.5979 | 0.5675 | 0.4445 | 0.5715 |
| | $\Delta E_0$ & $\Delta R_{InterD}$ | 0.704 | 0.4608 | | | |
| | $\Delta E_0$ & $\Delta R_{Ru\text{-}\mathrm{Mn}}$ | - | | | | 0.5532 |
| | $\Delta R_{Ru\text{-}\mathrm{O}}$ & $\Delta R_{\mathrm{Dimer}}$ | 0.4114 | 0.5555 | 0.5323 | 0.4138 | 0.5276 |
| | $\Delta R_{Ru\text{-}\mathrm{O}}$ & $\Delta R_{\mathrm{InterD}}$ | 0.5537 | | | | |
| | $\Delta R_{Ru\text{-}\mathrm{O}}$ & $\Delta R_{Ru\text{-}\mathrm{Mn}}$ | - | | | | 0.5027 |
| | $\Delta R_{\mathrm{Dimer}}$ & $\Delta R_{InterD}$ | | | | | -0.4939 |
| | $\Delta R_{\mathrm{Dimer}}$ & $\Delta R_{Ru\text{-}\mathrm{Mn}}$ | - | 0.6851 | 0.6733 | 0.594 | 0.7087 |
| | $\Delta R_{InterD}$ & $\Delta R_{Ru\text{-}\mathrm{Mn}}$ | - | | -0.5301 | -0.6991 | -0.7739 |
| | $\Delta\sigma_{\mathrm{InterD}}$ & $\Delta R_{Ru\text{-}\mathrm{Mn}}$ | - | 0.7281 | 0.6892 | | |

* There is no fitting parameter $\Delta R_{\mathrm{Mn}}$ and $\Delta\sigma_{\mathrm{Mn}}$ in the case of x = 0.

* Vacant or unmentioned correlations are all under 0.4.

*$E_0$ = 22123.4 eV.